\documentclass[doublecol]{epl2} 
\usepackage{hyperref}
\usepackage{epsfig}

\newcommand{\beq}{\begin{equation}}
\newcommand{\eeq}{\end{equation}}
\newcommand{\beqa}{\begin{eqnarray}}
\newcommand{\eeqa}{\end{eqnarray}}
\newcommand{\ba}{\begin{array}}
\newcommand{\ea}{\end{array}} 

\title{Supersonic and subsonic shock waves in the unitary Fermi gas} 
\author{Luca Salasnich}
\institute{Dipartimento di Fisica ``Galileo Galilei'' and CNISM, 
Universit\`a di Padova, Via Marzolo 8, 35131 Padova, Italy} 

\abstract{We investigate shock waves in the unitary Fermi 
gas by using the zero-temperature equations of 
superfluid hydrodynamics. We obtain analytical solutions for the 
dynamics of a localized perturbation of the uniform gas. 
These supersonic bright and subsonic dark solutions produce,  
after a transient time, an extremely large (divergent) density gradient: 
the shock. We calculate the time of formation of the shock and also simulate 
the space-time behavior of the waves by solving generalized 
hydrodynamic equations, which include a reliable dispersive regularization 
of the shock. We find that the shock spreads into wave ripples 
whose properties crucially depend on the chosen initial configuration.}


\begin{document}

\maketitle

One of the basic problems in physics is how density perturbations 
propagate through a material \cite{landau,whitham}. 
In addition to the well-known 
sound waves, there are shock waves characterized by 
an abrupt change in the density of the medium \cite{landau,whitham}. 
Shock waves are ubiquitous and 
have been studied in many different physical systems \cite{landau,whitham}. 
Ten years ago shock waves have been experimentally observed 
also in atomic Bose-Einstein condensates (BECs) 
\cite{hau,cornell,hoefer,davis}, and theoretically 
investigated in various BEC configurations 
\cite{zak,damski,gammal,perez,damski2,muga,noi,ablowitz}. 
Very recently the observation of nonlinear hydrodynamic waves 
has been reported in the collision between two strongly 
interacting Fermi gas clouds of $^6$Li atoms \cite{thomas}. 
The experiment shows the formation of density 
gradients, which are nicely reproduced by hydrodynamic equations 
with a phenomenological viscous term \cite{thomas}. 
Nevertheless, the role of dissipation is questionable 
\cite{new-bulgac} since the {ultracold} 
unitary Fermi gas is noted as an example 
of an almost perfect fluid \cite{sempre-thomas}. 
Indeed in Ref. \cite{new-bulgac} it has been shown,  
{by solving zero-temperature time-dependent 
Bogolibov-de Gennes equations}, that the viscous 
term is not necessary to {reproduce the experimental 
results of Ref. \cite{thomas}.} 

Here we investigate the formation 
and dynamics of shock waves in the dilute and ultracold unitary 
(divergent inter-atomic scattering length)
Fermi gas by using the zero-temperature equations of 
superfluid hydrodynamics \cite{landau}. 
At zero temperature fermionic superfluids in the BCS-BEC crossover can be 
modelled by hydrodynamic equations \cite{pitaevskii} 
and their generalizations with a gradient 
term \cite{luca12} that induces a dispersive regularization of the 
shock. In this paper we obtain analytical solutions for the 
dynamics of a localized perturbation of the uniform gas. 
We calculate the supersonic (or subsonic) velocity of propagation 
of these bright (or dark) perturbations. 
We show that bright perturbations 
evolve towards a shock-wave front, while dark perturbations 
produce the shock in their back, and 
we calculate the period $T_s$ of formation of the shock. In addition, 
we study the space-time behavior of the shock waves 
beyond this characteristic time $T_s$ by including a reliable quantum 
correction in the hydrodynamic equations \cite{luca12}. 
In fact, according to the two-fluid model 
of Landau \cite{landau}, the viscous term acts only on the normal 
component of the fluid, and at zero temperature the normal component 
is zero. Morover, recent theoretical microscopic calculations 
\cite{new-levin} suggest that 
the viscosity of the unitary Fermi gas is extremely small 
at very low temperatures because the transverse current does not couple 
to collective modes. 
By solving numerically these generalized equations of the unitary 
Fermi gas we show that the shock-wave front spreads into wave ripples 
whose properties crucially depend on the ``brightness'' (bright or dark) 
of the chosen initial configuration. 
{We expect that our results are reliable when the normal 
density (with its viscous term) is quite small, i.e. 
for a temperature $T$ much smaller than the critical temperature $T_c$ 
of the normal-superfluid transition. For the unitary Fermi gas 
$T_c\simeq 0.2\ T_F$, where $T_F=\epsilon_F/k_B$ 
is the Fermi temperature with $k_B$ the Boltzmann constant,  
$\epsilon_F=\hbar^2(3\pi^2 n)^{2/3}/(2m)$ 
the bulk Fermi energy of the ideal Fermi gas, 
$n$ the total density, and $m$ the mass of each atomic fermion. 
According to Ref. \cite{luca-termo} for $T<0.1\ T_F$ the normal fraction 
is below $10\%$ and, in practice, in this range of temperatures 
our approach is fully justified.}

At zero temperature the low-energy collective dynamics of a 
Fermi superfluid of neutral and dilute atoms 
at unitarity can be described by the equations of 
irrotational and inviscid hydrodynamics 
\beqa
{\partial\over \partial t} n  + \nabla \cdot (n {\bf v}) = 0 \; , 
\label{hy1}
\\
m {\partial\over \partial t} {\bf v} + \nabla 
\Big[ {m\over 2} v^2 + 
\mu(n) + U({\bf r}) \Big] = {\bf 0} \; ,   
\label{hy2} 
\eeqa
where $n({\bf r},t)$ is the total density of the superfluid, 
${\bf v}({\bf r},t)$ is its velocity field, and  
\beq 
\mu(n)= \xi {\hbar^2 \over 2m} (3\pi^2)^{2/3} n^{2/3} 
\label{chemical}
\eeq
is the bulk chemical potential of the system, with $\xi \simeq 0.4$ 
a universal parameter \cite{pitaevskii}. Here we are supposing 
a balanced system, namely the same number of fermions in the 
two components of the spin ($\sigma=\uparrow,\downarrow$).   
The term $U({\bf r})$ models the external potential 
which traps the atoms. 

Let us consider the unitary Fermi gas 
with constant density $\bar{n}$ with $U({\bf r})=0$. 
{Experimentally this configuration can be obtained 
with a very large square-well potential (or a similar external trapping), 
such that in the model one can effectively impose 
periodic boundary conditions instead of the vanishing ones.} 
A density variation along the $z$ axis with respect to the uniform 
configuration $\bar{n}$ can be experimentally created 
by using a blue-detuned (bright perturbation) 
or a red-detuned (dark perturbation) laser beam \cite{detuned}. 
In practice, we perform the following factorization 
\beq 
n({\bf r})= n_{\bot}(x,y) \, n_{\parallel}(z) \; , 
\eeq
by imposing also 
\beqa 
n_{\bot}(x,y) = \bar{n}_{\bot}
\\
n_{\parallel}(z) = \bar{n}_{\parallel} \; \rho(z) 
\eeqa
such that  
\beq 
n({\bf r}) = \bar{n} \, \rho(z) 
\label{ansatz}
\eeq
where $\bar{n}=\bar{n}_{\bot} \bar{n}_{\parallel}$, and 
$\rho(z,t)$ is the relative density, i.e. the 
localized axial modification with respect to the uniform 
density $\bar{n}$. {We impose periodic boundary 
conditions along the $z$ axis, namely $\rho(z=L_z,t)=\rho(z=-L_z,t)$, 
with $2L_z$ the axial-domain length.} 
We set ${\bf v}({\bf r},t)=(0,0,v(z,t))$ with $v(z,t)$ 
the velocity field {such that $v(z=L_z,t)=v(z=-L_z,t)$.} 
{Moreover we impose that the initial localized 
wave packet satisfies the boundary 
conditions $\rho(z=\pm L_z,t=0)=1$ and $v(z=\pm L_z,t=0)=0$.} 
Because the dimensional reduction is done 
assuming the uniformess in $x$,$y$ directions, we shall 
consider the propagation of a plane wave along the $z$ axis. 

Inserting Eq. (\ref{ansatz}) into Eqs. (\ref{hy1}) and (\ref{hy2}) 
one finds the 1D hydrodynamic equations for 
the axial dynamics of the superfluid, given by  
\beqa 
{\dot \rho} + v \rho' + v' \rho = 0 \; , 
\label{uf1}
\\
{\dot v} + v v' +  {c_{ls}(\rho)^2\over \rho} \rho' = 0 \; ,    
\label{uf2}
\eeqa 
where dots denote time derivatives, primes space derivatives,  
and 
\beq 
c_{ls}(\rho) = c_s \rho^{1/3} 
\label{nice-cs}
\eeq
is the local sound velocity, with 
$c_s=c_{ls}(1)=\sqrt{\xi/3}v_F$ the bulk sound velocity, 
$v_F=\sqrt{2\epsilon_F/m}$ is bulk Fermi velocity 
and $\epsilon_F={\hbar^2\over 2m }(3\pi^2 \bar{n})^{2/3}$ 
the bulk Fermi energy.

The bulk sound velocity $c_s$ is the speed of 
propagation of a small perturbation $\tilde{\rho}(z,t)$ 
with respect to the uniform superfluid of density $\bar{n}$. 
In fact, setting 
$\rho(z,t)=1+ \tilde{\rho}(z,t)$, with $\tilde{\rho}(z,t)\ll 1$ and 
$v(z,t)$ of the same order of $\tilde{\rho}(z,t)$, 
from the linearization of Eqs. (\ref{uf1}) and (\ref{uf2}) we get 
the familiar linear wave equation 
\beq 
\left( {\partial^2\over \partial t^2} - c_s^2 
{\partial^2 \over \partial z^2} \right) \tilde{\rho}(z,t) = 0 \; ,    
\eeq 
for $\tilde{\rho}(z,t)$ and a similar equation for $v(z,t)$. 
Modelling the initial perturbation with a Gaussian shape, i.e. 
\beq 
\tilde{\rho}(z,0) = 2\eta \ e^{-z^2/(2\sigma^2)} \; , 
\label{initial-rho} 
\eeq
one finds \cite{landau,whitham} from the linearized equations 
\beq 
\tilde{\rho}(z,t) = \eta \ e^{-(z-c_st)^2/(2\sigma^2)} + 
\eta \ e^{-(z+c_st)^2/(2\sigma^2)} \; ,  
\label{inso}
\eeq
{with initial condition ${\dot {\tilde \rho}}(z,t=0)=0$.} 
Thus, for the conservation of the linear momentum, 
the initial wave packet splits into two waves 
travelling in opposite directions with the speed of sound $c_s$. 
Obviously Eq. (\ref{inso}) is reliable only if $|\eta| \ll 1$. 
As expected, a small (infinitesimal) perturbation 
gives rise to sound waves. 

We can find wave solutions of Eqs. (\ref{uf1}) and (\ref{uf2}) 
with a generic initial density profile 
by following the approach described by Landau and  Lifshitz \cite{landau}.  
By supposing that the velocity $v$ depends explicitly 
on the density $\rho$, i.e. $v=v(\rho(z,t))$, one has ${\dot v}=
{d v \over d \rho} {\dot \rho}$,  
$v'={d v \over d \rho} \rho'$. 
We now impose that the two hydrodynamic equations reduce to 
the same hyperbolic equation 
\beq 
{\dot \rho} + c(\rho) \rho' = 0 \; , 
\label{pippo}
\eeq 
where 
\beq 
c(\rho ) = v(\rho) + {d v \over d \rho}\rho 
= v(\rho) + {c_{ls}(\rho)^2 \over \rho} 
\left({d v \over d \rho} \right)^{-1} \; .   
\label{cc2}
\eeq 
It is quite easy to verify that, given a initial condition 
$F(z)=\rho(z,t=0)$ for the density profile, 
the time-dependent solution $\rho(z,t)$ 
of the hyperbolic equation (\ref{pippo}) satisfies the following implicit, 
but algebraic, equation: 
\beq 
\rho(z,t) = F(z - c(\rho(z,t)) t) \; . 
\label{exact-rho}
\eeq 

To determine the local velocity of propagation 
$c(\rho)$, which is not equal to the sound velocity $c_{ls}(\rho)$, 
we observe that from Eqs. (\ref{cc2}) we get 
\beq 
{d v \over d \rho} = \pm {c_{ls}(\rho) \over \rho}   \; .  
\eeq 
After separation of variables, and imposing that at infinity the 
density is equal to one and the velocity field is zero,  
we finally get  
\beq 
v(\rho ) = \pm 3 c_s (\rho^{1/3} - 1) \; .    
\label{nice-v} 
\eeq 
The velocity $c(\rho)$ follows directly from the velocity 
$v(\rho)$ by using Eqs. (\ref{cc2}). One finds 
$c(\rho ) = v(\rho) \pm c_{ls}(\rho)$, namely 
\beq 
c(\rho ) = \pm c_s \left( 4 \rho^{1/3} - 3 \right) 
\; . 
\label{nice-c}
\eeq 
In conclusion, we have found that the density field $\rho(z,t)$ 
satisfies the implicit algebraic equation (\ref{exact-rho}) 
with $c(\rho)$ given by Eq. (\ref{nice-c}). 
Note that a similar result, 
but with a very different local velocity, has been 
obtained by Damski \cite{damski} for the 1D BEC. 

\begin{figure}
\centerline{\epsfig{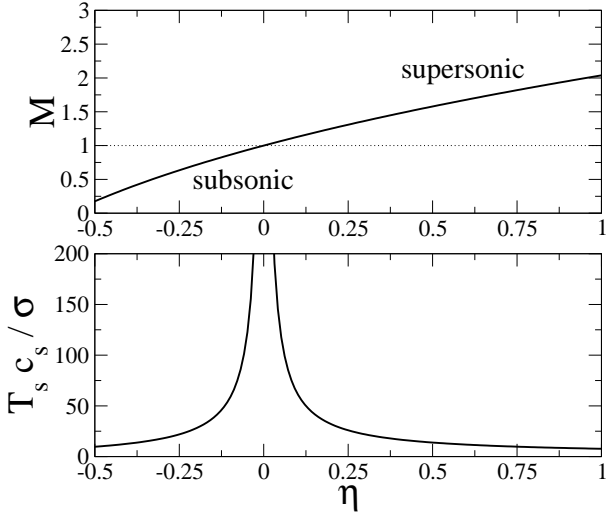}}
\small 
\caption{Properties of the shock waves. 
Upper panel: Mach number $M$ as a function of the amplitude $\eta$ 
of the perturbation (solid line). For $M<1$ there are subsonic and 
dark ($-1/2\leq \eta<0$) waves, while for $M>1$ 
there are supersonic and bright ($\eta>0$) waves. 
The dotted line separates the two regimes. 
Lower panel: period $T_s$ of formation (breaking time) 
of the shock-wave front 
as a function of the amplitude $\eta$ of the perturbation. 
$T_s$ is in units of $\sigma/c_s$, where $\sigma$ is the width 
of the perturbation and $c_s=\sqrt{\xi/3}v_F$ 
is the bulk speed of sound, with $v_F$ the Fermi velocity.} 
\label{fig1}
\end{figure} 

Eqs. (\ref{exact-rho}) and (\ref{nice-c}) contain the dynamics 
of the two waves propagating to the left and to the right 
with initial condition (\ref{initial-rho}). 
Some properties characterizing the dynamics can 
be extracted from these equations. First of all the two 
travelling waves have symmetric shapes during the time evolution. 
In addition, both amplitude and velocity 
of the extrema (maxima or minima, depending on the sign of $\eta$) 
of the two waves are practically constant during time evolution. 
In particular, the amplitude of the extrema is given by 
$A(\eta) = 1+\eta$ while the velocity of the extrema reads 
\beq 
V(\eta)= c(1+\eta) = \pm c_s 
\left(4(1+\eta)^{1/3}-3 \right)\; . 
\label{nice-vmax}
\eeq
Notice that taking $\eta =0$ the velocity of the impulse 
extrema reduces to the sound velocity: $V(0)=c(1)=c_s=\sqrt{\xi/3}v_F$. 
Moreover, bright perturbations ($\eta >0$) move 
faster than dark ones ($\eta < 0$), and the Mach number 
$M=V(\eta)/V(0)$ of these perturbations 
in the unitary Fermi gas is simply 
\beq 
M = 4(1+\eta)^{1/3}-3 \; .   
\eeq 
For $M>1$, which means $\eta>0$ (bright perturbation),  
one has supersonic waves, 
while for $0\leq M<1$, which means $\eta<0$ (dark perturbation),  
one has subsonic waves.  In the upper panel 
of Fig. \ref{fig1} we plot the Mach number $M$ as function of the 
amplitude $\eta$ of the perturbation. Note that since $2\eta$ 
is the amplitude of the initial condition, 
see Eq. (\ref{initial-rho}), the region $\eta \leq -1/2$ is unphysical. 

Let us consider a bright perturbation ($\eta>0$) moving to the right. 
The speed of impulse maximum $V(\eta)$ 
is bigger than the speed of its tails $V(0)$. 
As a result the impulse self-steepens in the direction 
of propagation and a shock wave front 
takes place. The breaking-time $T_s$ required for such a process can 
be estimated as follows: the shock wave front appears 
when the distance difference traveled by lower and upper impulse 
parts is equal to the impulse half-width $2\sigma$, 
namely $[V(\eta)-V(0)] T_s = 2 \sigma $. 
This, by using Eqs. (\ref{nice-c}) and Eq. (\ref{nice-cs}), 
gives  
\beq 
T_s = {\sigma \over 2c_s ((1+\eta)^{1/3}-1) } 
\; .  
\label{nice-ts}
\eeq 
In the case of a dark perturbation ($\eta <0$) the tails  
of the wave packet move faster than the impulse minimum. 
The shock appears in the back of the travelling wave,  
and the period of shock formation is simply 
$T_s =2 \sigma/(V(0) - V(\eta))$. In the lower panel 
of Fig. \ref{fig1} we plot the period $T_s$ as a function 
of the amplitude $\eta$ of the perturbation. The figure shows that 
as $\eta$ goes to zero the period $T_s$ goes to infinity; 
in fact, in this limit the shock wave reduces to a 
sonic wave (sound wave) which does not produce a shock. 

After the formation of the shock 
Eqs. (\ref{hy1}) and (\ref{hy2}) are not reliable 
because their exact solutions given of Eqs. (\ref{exact-rho}) 
and (\ref{nice-c}) are no more single-valued. To overcome this difficulty 
we include a gradient quantum term 
in the hydrodynamic equations, which become 
\beqa
{\partial\over \partial t} n  + \nabla \cdot (n {\bf v}) = 0 \; , 
\label{hy1g}
\\
m {\partial\over \partial t} {\bf v} + \nabla 
\Big[ {m\over 2} v^2 
- \lambda {\hbar^2\over 2m} {\nabla^2 \sqrt{n}\over \sqrt{n}} + 
\mu(n) + U({\bf r}) \Big] = {\bf 0} \; ,   
\label{hy2g} 
\eeqa
We stress that at zero temperature the simplest regularization 
process of the shock is a purely dispersive quantum gradient term. 
Historically, the gradient term with $\lambda$ 
was introduced  by von Weizs\"acker to treat surface effects 
in nuclei \cite{von}. This approach has been 
adopted for quantum hydrodynamics of electrons by 
March and Tosi \cite{tosi}, and also by Zaremba and 
Tso \cite{tso}. In the study of the BCS-BEC crossover the gradient 
term has been considered by various authors \cite{various}. 
The choice of the parameter $\lambda$ in Eqs. (\ref{hy1g}) and (\ref{hy2g}) 
is still debated, 
here we choose $\lambda=1/4$, a value which gives good agreement 
with Monte Carlo calculations at zero and finite temperature 
(for details see \cite{luca12,luca-termo}). Moreover we set $\xi=0.4$. 

We expect that Eqs. (\ref{hy1g}) 
and (\ref{hy2g}) are reliable to study the long-time dynamics 
of shock waves in the ultracold unitary Fermi gas. 
It is well known that, according to the two-fluid model and 
the Landau's criterion of superfluidity \cite{landau},  
above a critical temperature $v_c$ 
a normal component with a dissipative term appears 
in the fluid \cite{landau}. As discussed in \cite{combescot}, 
for the unitary Fermi gas one has $v_c \simeq c_s$. 
Nevertheless, at ultrcold temperature the normal component is 
negligible and also the shear viscosity \cite{sempre-thomas,new-levin}. 
For these reasons at zero temperature the shock waves are 
dispersive and not dissipative \cite{new-bulgac}. 

Eqs. (\ref{hy1g}) and (\ref{hy2g}) can be formally written 
(for any value of $\lambda$, also $\lambda=0$) in terms of a 
Galilei-invariant nonlinear Schr\"odinger equation \cite{luca12}. 
Setting $U({\bf r})=0$ 
and using Eqs. (\ref{chemical}) and (\ref{ansatz}) 
we easily get from 
Eqs. (\ref{hy1g}) and (\ref{hy2g}) a 1D nonlinear Sch\"odinger equation. 
We solve this equation by using a Crank-Nicolson 
finite-difference predictor-corrector algorithm \cite{sala-numerics} 
with the initial condition given by Eq. (\ref{initial-rho}) 
and $v(z,t=0)=0$. In fact, as also shown by Damski \cite{damski}, 
we have verified that the initial velocity field $v(\rho(z,t=0))$ and  
$v(z,t=0) = 0$ give practically the same time evolution. 

\begin{figure}
\centerline{\epsfig{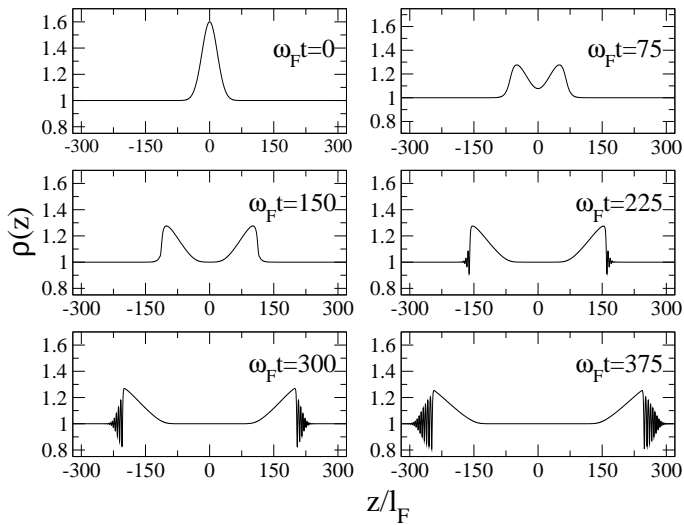}}
\small 
\caption{Time evolution of supersonic shock waves. 
Initial condition with $\sigma/l_F=18$ and $\eta=0.3$. 
The curves give the relative density profile $\rho(z)$ 
at subsequent frames, where 
$l_F=\sqrt{\hbar^2/(m\epsilon_F)}$ is the Fermi length 
and $\omega_F=\epsilon_F/\hbar$ is the Fermi frequency.} 
\label{fig2}
\end{figure} 

In Fig. \ref{fig2} we plot the time evolution of supersonic shock waves 
obtained with $\sigma/l_F=18$ and $\eta=0.3$, with 
$\sqrt{\hbar^2/(m\epsilon_F)}$ the Fermi length of the bulk system. 
The figure displays the density profile $\rho(z)$ 
at subsequent times. Note the splitting 
on the initial bright wave packet into two bright travelling waves moving 
in opposite directions. As previously discussed, 
there is a deformation of the two waves with the formation 
of a quasi-horizontal shock-wave front. Eventually, 
this front spreads into wave ripples. 
There is no qualitative difference with respect 
to a Bose-Einstein condensate 
\cite{damski} in the physical manifestation of supersonic shock waves 
in the zero-temperature unitary Fermi gas. 
Nevertheless, due to the very different equation 
of state, there are large quantitative differences. 
Our numerical analysis confirms that the breaking time $T_s$ decreases 
by increasing the amplitude $\eta$, while the velocity $V$ of the maxima 
of the travelling waves increases by increasing $\eta$. 
There is a good agreement between our analytical formulas, 
Eqs. (\ref{nice-vmax}) and (\ref{nice-ts}), and simulations: 
the relative difference is within $5\%$ for the velocity $V$ 
and within $20\%$ for the breaking time $T_s$. 

\begin{figure}
\centerline{\epsfig{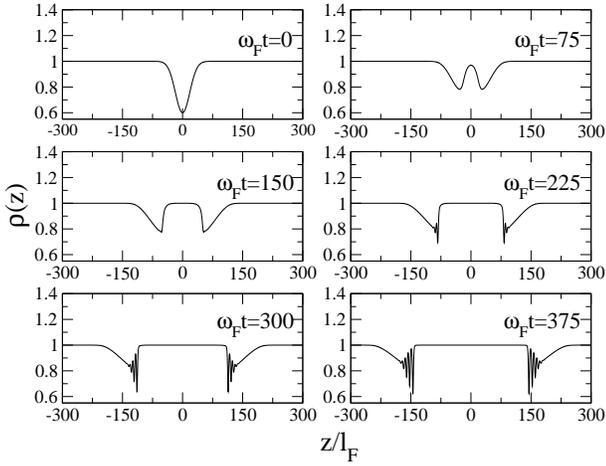}}
\small 
\caption{Time evolution of subsonic shock waves. 
Initial condition with $\sigma/l_F=18$ and $\eta=-0.2$. 
The curves give the relative density profile $\rho(z)$ 
at subsequent frames, where 
$l_F=\sqrt{\hbar^2/(m\epsilon_F)}$ is the Fermi length 
and $\omega_F=\epsilon_F/\hbar$ is the Fermi frequency.} 
\label{fig3}
\end{figure}

In Fig. \ref{fig3} we plot the time evolution of subsonic shock waves 
obtained again with $\sigma/l_F=18$ and 
$\eta=-0.2$. Also in this case the figure shows the splitting 
on the initial dark wave packet into two dark travelling waves moving 
in opposite directions. But here, as expected, 
the quasi-horizontal shock 
appears in the back side of the travelling waves. 
Our simulations show that for $\eta<0$ 
the wave ripples which appear at the breaking time $T_s$ 
are always dark, i.e. they never exceed the bulk density 
(compare wave ripples of Fig. \ref{fig2} with those of Fig. \ref{fig3}). 
Note that dark shock waves have been studied long 
ago \cite{mario} in a different physical context: the discrete 
nonlinear Schr\"odinger equation. In that case 
the wave ripples can exceed the bulk density, probably due to the 
discrete nature of the Schr\"odinger equation. 
Also for dark shock waves 
our analytical predictions on velocity $V$ 
of the minima and breaking time $T_s$ are 
quite accurate with respect to numerical findings 
(similar relative differences of runs with $\eta>0$). 

In conclusion, we have shown that at very low 
temperatures the unitary Fermi gas admits supersonic and subsonic 
shock waves, for which we have developed analytical and numerical 
results. Our predictions suggest a much cleaner method to 
produce shock waves with respect to the recent experiment \cite{thomas}
based on the collision of two $^6$Li atomic clouds. 
The shape of these waves changes during the time evolution 
giving rise to a shock-wave front at a characteristic 
breaking time. We have determined the Mach number of these travelling 
waves as a function of the perturbation amplitude, showing that 
supersonic bright and subsonic dark waves behave quite differently. 

The author thanks Sadahn Kumar Adhikari, Tilman Enss, 
{Boris Malomed}, Enzo Orlandini, Mario Salerno, and Flavio Toigo 
for useful suggestions and discussions.

\end{document}